%% file: CCNR.tex
\pdfoutput=1
\documentclass[prd,aps,superscriptaddress,amsmath,amssymb,twocolumn,nobibnotes,nofootinbib,10pt]{revtex4-2}

\usepackage{braket,verbatim,bm,bbold,amsmath,amssymb,epsfig,float,xcolor}
\definecolor{redCB}{RGB}{150,50,60}  
\definecolor{dteal}{RGB}{7,94,84}  
\definecolor{blueCB}{RGB}{108,144,170}
\usepackage[colorlinks=true,    	
			pdfstartview=FitV,
            bookmarks=false,
            citecolor=dteal,
            linkcolor=dteal,
           linkcolor=dteal,
            hyperfootnotes=true,
            linktoc=page,
            urlcolor=dteal]{hyperref}
\usepackage{soul}

\newcommand{\nc}{\newcommand}
\nc{\ir}{\mathrm{i}}
\nc{\dd}{\mathrm{d}} 
\nc{\eE}{\mathrm{e}}
\nc{\Tr}{\text{Tr}}
\nc{\id}{\mathbb{I}}
\nc{\Z}{\mathcal{Z}}
\nc{\E}{\mathcal{E}}
\nc{\F}{\mathcal{F}}
\nc{\Om}{\Omega}
\nc{\N}{\mathcal{N}}
\nc{\spp}{\hspace{1pt}}
\nc{\spm}{\hspace{-1pt}}
\nc{\lb}{\label} 
\nc{\nn}{\nonumber} 
\nc{\ra}{\rangle} 
\nc{\la}{\langle} 
\nc{\Nn}{\mathcal{N}} 
\nc{\sigb}{\boldsymbol{\sigma_{\text{bd}}}}
\nc{\HA}{\mathcal{H}_A}
\nc{\HB}{\mathcal{H}_B}
\nc{\HC}{\mathcal{H}_C}
\nc{\ta}{\tilde{a}}
\nc{\eps}{\epsilon}
\nc{\Q}{\mathcal{Q}}
\nc{\T}{\mathcal{T}}
\nc{\Oo}{\mathcal{O}}

\def\bea#1\eea{\begin{align}#1\end{align}}
\def\bes#1\ees{\begin{equation}\begin{split}#1\end{split}\end{equation}}

\begin{document}

\title{Reflected entropy and computable cross-norm negativity:\\Free theories and symmetry resolution }

\author{Cl\'ement Berthiere}
\email{clement.berthiere@umontreal.ca}
\affiliation{\it Centre de Recherches Math\'ematiques, Universit\'e de Montr\'eal, Montr\'eal, QC, H3C 3J7, Canada}
\affiliation{\it D\'epartement de Physique, Universit\'e de Montr\'eal, Montr\'eal, QC, H3C 3J7, Canada}

\author{Gilles Parez}
\email{gilles.parez@umontreal.ca}
\affiliation{\it Centre de Recherches Math\'ematiques, Universit\'e de Montr\'eal, Montr\'eal, QC, H3C 3J7, Canada}
\affiliation{\it D\'epartement de Physique, Universit\'e de Montr\'eal, Montr\'eal, QC, H3C 3J7, Canada}

\date{\today}

\begin{abstract}
We investigate a separability criterion based on the computable cross-norm (CCNR), and a related quantity called the CCNR negativity. We introduce a \textit{reflected} version of the CCNR negativity, and discuss its connection with other well-established entanglement-related quantities, namely the reflected entropy and the operator entanglement entropy. 
For free fermionic and bosonic theories, we derive exact formulas in terms of two-point correlation functions, which allow for systematic numerical investigations and, in principle, analytical treatments. For systems with a global $U(1)$ symmetry, we study the symmetry-resolved reflected entropy and CCNR negativity. We provide conformal field theory (CFT) results for the charged moments in the case of adjacent intervals, finding perfect agreement with the numerics. We observe an equipartition of reflected entropies and CCNR negativities, both for free-fermions and free-boson models. The first charge-dependent corrections are conjectured for fermions, and worked out from the CFT calculations for bosons. 
\end{abstract}

\maketitle
\makeatletter

\def\l@subsubsection#1#2{}
\makeatother

\tableofcontents

\section{Introduction}

A fundamental problem \cite{horodecki2009quantum,2009PhR...474....1G} in quantum information theory is the detection of entanglement for general quantum states. Since the pioneering work of Werner \cite{PhysRevA.40.4277} concerning the problem of separability of mixed states, several criteria have been proposed to decide whether a given quantum state is entangled or not. Among them stand out the positive partial transposition (PPT) criterion \cite{Peres:1996dw,Horodecki:1996nc} and the computable cross-norm or realignment (CCNR) criterion \cite{2003JPhA...36.5825R,2002quant.ph..5017C}, providing simple necessary conditions for separability. These two independent criteria are based on permutations of density matrix elements \cite{2002quant.ph..6008H,Liu:2022ehb}, and neither is stronger in general. 

Based on the PPT criterion, the PPT logarithmic negativity \cite{zyczkowski1998volume,Eisert:1998pz,Vidal:2002zz} is a well-established entanglement measure for bipartite mixed states, in particular in the context of quantum many-body systems \cite{Calabrese:2012ew,Calabrese:2012nk,2013PhRvA..88d2319C,2013PhRvA..88d2318L,Wen:2016snr,Angel-Ramelli:2020wfo}. Recently, a CCNR negativity based on the CCNR criterion has been discussed in conformal field theory (CFT) \cite{Yin:2022toc}, holography \cite{Milekhin:2022zsy} and topological systems \cite{Yin:2023jad}. 

Entanglement plays a prominent role in quantum many-body systems \cite{Amico:2007ag,Calabrese:2009qy,Laflorencie:2015eck}, particularly in relation to critical properties \cite{Vidal:2002rm,Calabrese:2004eu,2006PhRvL..96j0603L,Eisert:2008ur,2009JPhA...42X4009A,Bueno:2015rda,Fursaev:2016inw,Casini:2016fgb,Berthiere:2018ouo,Berthiere:2019lks} and topological order \cite{Kitaev:2005dm,2006PhRvL..96k0405L,2015arXiv150802595Z}.
In the context of quantum many-body systems with a global conserved charge, an important issue is to understand how each symmetry sector contributes to the total entanglement. This \textit{symmetry resolution} of entanglement is the object of intense research, both theoretical and experimental \cite{lr-14,GS18,equi-sierra,bons,exp-lukin, vitale2022symmetry,neven2021symmetry,Rath:2022qif}. Symmetry-resolved entanglement entropies have notably been investigated in the context of critical systems \cite{GS18,equi-sierra,bons,bons-20,barghathi2018renyi,barghathi2019operationally,mrc-20,eimd-21,ares2022multi,ares2022symmetry,di2022boundary}, integrable field theories and lattice models \cite{Murciano:2019wdl,Murciano:2020vgh,hc-20,hcc-21,capizzi2022symmetry,capizzi2022entanglement}, topological phases \cite{as-20,Topology,oblak2022equipartition,horvath2023charge,fossati2023symmetry,Topology2} and quantum many-body systems out of equilibrium \cite{fg-19,fg-21,PBC21, PBC21bis, PBC22,parez2022analytical,piroli2022thermodynamic,scopa2022exact,murciano2023symmetry}. In addition, other entanglement-related quantities possess a meaningful symmetry resolution \cite{cgs-18,mbc-21,gaur2023charge,cc-21,Rath:2022qif,murciano2022symmetry,parez2022symmetry,ares2023entanglement,di2023symmetry}. In this paper, we investigate the symmetry resolution of two related quantities, the CCNR negativity and reflected entropies, in the context of CFT and free theories.

This paper is organized as follows. We begin in Sec.~\ref{sec:quantities} with the definition of the (R\'enyi) CCNR negativity, and its \textit{reflected} generalization that we introduce. We discuss the relations between the Rényi CCNR negativities and two other important quantities, namely the reflected entropy and the operator entanglement entropy. In Sec.~\ref{Gauss_states}, we derive general formulas for the R\'enyi reflected entropies and CCNR negativities, valid for free fermionic and bosonic fields in arbitrary dimensions. The resulting expressions are fully determined in terms of the two-point correlation functions,
making them suitable for lattice calculations. We study the symmetry resolution of reflected entropy and CCNR negativity in Sec.~\ref{sec:sym-res}. We present CFT results for two adjacent regions, which we verify against numerical calculations for free fermions and bosons.
We conclude in Sec.~\ref{conclu} with a summary of our main results, and give an outlook on future study. Finally, some technical details can be found in the three appendices that complete this work.

\section{Quantities of interest}\lb{sec:quantities}

In this section we define the CCNR negativity, the reflected entropy and the operator entanglement entropy. We discuss various connections between these quantities and their important properties. 

\subsection{Definition of the CCNR negativity}
Let $\rho_{AB}$ be the density matrix of a bipartite system~$A~\cup~B$ in a general state. By definition, $\rho_{AB}$ is a positive semidefinite Hermitian operator with trace one. We assume that the total Hilbert space factorizes as the product of local Hilbert spaces $\HA\otimes\HB$, and define $\{|a\rangle\}$ and $\{|b\rangle\}$ to be orthonormal basis states of $\HA$ and $\HB$, respectively. With this choice of bases, the density matrix reads 
%
\begin{equation}
\rho_{AB}=\sum_{a,a'}\sum_{b,b'} \langle a b| \rho_{AB}|a'b'\rangle\,|ab \rangle \langle a'b'| \spp,
\end{equation}
where $|ab\rangle \equiv |a\rangle \otimes |b\rangle$. We introduce the realignment matrix of $\rho_{AB}$, denoted $R\equiv R(\rho_{AB})$ as
\bea
 R=\sum_{a,a'}\sum_{b,b'} \langle ab| \rho_{AB}|a'b'\rangle\,|aa'\rangle \langle bb'| \spp
\eea
by swapping basis elements $|b\ra$ and $|a'\ra$ of $\rho_{AB}$. By definition, $R$ is not necessarily a square matrix, though the product $RR^\dagger$ is. The latter operator acts on $\HA \otimes \HA$,
\begin{equation}\lb{RRdag}
    RR^\dagger = \sum_{a,a'}\sum_{\ta,\ta'} \Tr_{\HB}\Big(\spm\langle a| \rho_{AB} |a'\rangle \langle \ta' | \rho_{AB} | \ta \rangle \spm\Big)\,|aa'\rangle \langle \ta \ta'| \spp.
\end{equation}
For separable states, \mbox{$||R||=\Tr\big( \sqrt{RR^\dagger}\big)\leqslant 1$}, which constitutes the CCNR separability criterion \cite{2003JPhA...36.5825R,2002quant.ph..5017C}. A state is thus guaranteed to be entangled if $||R||> 1$.

Analogously to the PPT logarithmic negativity, the CCNR negativity is defined as
\begin{equation}\lb{CCNR_neg}
    \E = \log \Tr \big( \sqrt{RR^\dagger}\big)\spp.
\end{equation}
The trace norm $||R||$ is challenging to compute in practice. Fortunately, a replica formulation of \eqref{CCNR_neg} can be implemented \cite{Yin:2022toc}.
The replica method relates the trace norm of $R$ to the moments of $RR^\dagger$,
\bea
Z_n = \Tr \big( RR^\dagger \big)^n \spp,
\eea
where $||R||$ is obtained by analytic continuation from integer values $n$ to $1/2$.
A R\'enyi generalization of \eqref{CCNR_neg} can be defined as
\bea
\E_n = \log\Tr \big( RR^\dagger \big)^n\spp,
\eea
such that $\E=\lim_{n\to1/2}\E_n$. For $n=1$ we retrieve the purity $\Tr\big( RR^\dagger\big) = \Tr\spp\rho_{AB}^2$ such that $\E_1=-S_2(\rho_{AB})$ where $S_n(\rho)\spm=\spm(1-n)^{-1}\spm\log\Tr\spp\rho^n$ is the R\'enyi entropy of~$\rho$.

For pure states, it is straightforward to show that
\bea\label{eq:ccnr_pure}
\quad \E_n=2\log\Tr\spp\rho_A^n = 2(1-n)S_n(\rho_A)\spp,   
\eea
where $\rho_A=\Tr_{\HB}\rho_{AB}$, such that the CCNR negativity is the R\'enyi entropy of order $1/2$, $\E=S_{1/2}(\rho_A)$. For simple separable states, $\rho_{AB}=\sum_k p_k \rho_A^{(k)}\otimes\rho_B^{(k)}$ where $\rho_B^{(k)}$ are projectors with orthogonal support, we have
\bea
\E_n = \log\Big(\sum_k p_k^{2n}\Big)\spp,
\eea
which yields a vanishing CCNR negativity for $n=1/2$ since $\sum_k p_k=1$, in accordance with the criterion $\E \leqslant 0$ for separable states.

\subsection{A generalization of the realignment matrix}
We may generalize the realignment matrix of $\rho_{AB}$ to that of $\rho_{AB}^{m/2}$ with $m\in\mathbb{Z}^+$ as
\bea\lb{Rm}
 R_m=\sum_{a,a'}\sum_{b,b'} \langle ab| \rho_{AB}^{m/2}|a'b'\rangle\,|aa'\rangle \langle bb'| \spp.
\eea
For $m=2$ it reduces to $R$. 
Similarly to the special case $m=2$ discussed above, we introduce the moments
\bea
\label{eq:ccnr_moments}
Z_{m,n} = \Tr \big( R_mR_m^\dagger \big)^n \spp,
\eea
and define the $(m,n)$--R\'enyi CCNR negativity as
\bea\lb{CCNRneg_def}
\E_{m,n} = \log\Tr \big( R_mR_m^\dagger \big)^n\spp,
\eea
such that $\E_{2,n}\equiv\E_n$.
For pure states, it still holds that $\E_{m,n}= 2(1-n)S_n(\rho_A)$, irrespective of $m$. Moreover, we have the relation $\Tr\big( R_mR_m^\dagger\big) = \Tr\spp\rho_{AB}^m$ such that $\E_{m,1}=(1-m)S_m(\rho_{AB})$.

\subsection{Relation to reflected entropy and operator entanglement entropy}
In this subsection, we discuss the relations between three information theoretic quantities of recent interest: the CCNR negativity, the reflected entropy and the operator entanglement entropy. 

The reflected entropy \cite{Dutta:2019gen} has attracted much attention \cite{Jeong:2019xdr,Bueno:2020vnx,Li:2020ceg,Berthiere:2020ihq,Bueno:2020fle,Akers:2022max,Chen:2022fte,Vasli:2022kfu,Lu:2022cgq,Afrasiar:2022fid,Sohal:2023hst,Basak:2023uix,Berthiere:2023bwn} since its introduction as a correlation measure for mixed states in the holographic context. However, it has recently been shown that the reflected entropy is not a measure of physical correlations in general, since it is not monotonically nonincreasing under partial trace \cite{Hayden:2023yij}. Nevertheless, this quantity has a meaningful relationship to entanglement, in particular with tripartite entanglement \cite{Akers:2019gcv,Hayden:2021gno} (see  also Appendix~\ref{apdx:markov}), and with mutual information \cite{Bueno:2020fle,Camargo:2021aiq}.
To define the reflected entropy, we consider a purification of $\rho_{AB}^m$ with \mbox{$m\in2\mathbb{Z}^+$}, denoted $|\Omega_m\ra$, such that  $\Tr_{\HA\otimes \HB}\big(|\Omega_m\ra\la\Omega_m|\big)=\frac{1}{\Tr\rho_{AB}^m}\rho_{AB}^m$. It can be constructed using the Choi-Jamiołkowski isomorphism \cite{Jamiokowski1972LinearTW,Choi1975CompletelyPL} as
\begin{equation}\lb{Choi}
    |\Omega_m\ra=\frac{1}{\sqrt{\Tr \rho_{AB}^m}}\sum_{a,a'}\sum_{b,b'} \langle ab| \rho_{AB}^{m/2}|a'b'\rangle\,|ab\rangle \otimes | a'b'\rangle\spp,
\end{equation}
by doubling the original Hilbert space. 
A replica formulation of the reflected entropy gives a practical handle for computations. This replica trick involves two replica indices, $m$ and $n$. The latter represents the usual R\'enyi index while the former is related to the purification. One then defines the reflected density matrix $\rho_{AA}^{(m)}=\Tr_{\HB\otimes \HB}\big(|\Omega_m\ra\la\Omega_m|\big)$, and introduces the $(m,n)$--R\'enyi reflected entropy as 
\bea\lb{REmn}
S^{R}_{m,n}=\frac{1}{1-n}\log\Tr\big(\rho_{AA}^{(m)}\big)^n\,.
\eea
The $n$--R\'enyi reflected entropy is recovered by analytic continuation $m\rightarrow1$, and the reflected entropy by further taking the limit $n\rightarrow 1$.

The reflected density matrix $\rho_{AA}^{(m)}$, constructed using the Choi-Jamiołkowski isomorphism as described above, can also be defined using the realignment matrix of $\rho_{AB}^{m/2}$ (see \eqref{Rm}) as
\bea\label{eq:rhoAAmRR}
\rho_{AA}^{(m)} =\frac{1}{\Tr \rho_{AB}^m}R_mR_m^\dagger \spp.
\eea
Comparing the $(m,n)$--R\'enyi CCNR negativity in \eqref{CCNRneg_def} with the $(m,n)$--R\'enyi reflected entropy in \eqref{REmn}, we have
\bea\lb{CCNR_RE}
\E_{m,n} &= (1-n)S^R_{m,n} + n(1-m)S_m(\rho_{AB}) \,,
\eea
such that setting $m=2$ we get the R\'enyi CCNR negativity $\E_n$. The $(m,n)$--R\'enyi CCNR negativity can be viewed as the unnormalized $(m,n)$--R\'enyi reflected entropy. With expression \eqref{CCNR_RE}, it can be checked that for topological systems, the results of \cite{Yin:2023jad} on CCNR negativity match earlier work on reflected entropy \cite{Berthiere:2020ihq}.

The operator entanglement entropy of a density matrix \cite{Zanardi:2001zza,2007PhRvA..76c2316P,Dubail:2016xht,Rath:2022qif}, denoted $E_n$ (for its R\'enyi generalization), is the (R\'enyi) Shannon entropy of the squared probability distribution values of the operator Schmidt decomposition of that density matrix. It is direct to realize that the reflected entropy is the operator entanglement entropy of $\sqrt{\rho_{AB}}$, that is
\bea\label{eq:En_SnR}
E_n = \frac{1}{1-n}\log\Tr\big(\rho_{AA}^{(2)}\big)^n = S^R_{2,n}\spp,
\eea
implying that $\E_{n} = (1-n)E_n - nS_2(\rho_{AB})$.

\subsection{CCNR negativity is not a correlation measure}

Here we point out that, contrary to what has been commonly stated in the recent literature, the CCNR negativity, similarly to the reflected entropy \cite{Hayden:2023yij}, is not a correlation measure. 

Beside the nonnegativeness required of a bipartite correlation measure $\mathcal{C}(A:B)$, which is clearly not satisfied by the (R\'enyi) CCNR negativity, one fundamental requirement is that correlation cannot increase under local operations. This means that a correlation measure should not increase under the local discarding of information, i.e.~the following inequality must hold
\bea\lb{monotone_ineq}
\mathcal{C}(A:B\cup C) \geqslant \mathcal{C}(A:B)\spp.
\eea
A quantity satisfying this inequality is said to be monotonically nonincreasing under partial trace.
We prove in Appendix\;\ref{app:nonm} that the R\'enyi CCNR negativity is \textit{not} monotonically nonincreasing under partial trace in the range $n\in (0,1)\cup(1,\infty)$, by providing counterexamples to inequality \eqref{monotone_ineq}.
As a byproduct, we prove that the $(m,n)$--R\'enyi reflected entropy $S^R_{m,n}$, which reduces to the R\'enyi operator entanglement entropy $E_n$ for $m=2$, is not monotonically nonincreasing under partial trace in the range $m>0$ and $n\in (0,2)$.
We thus conclude that the CCNR negativity and operator entanglement entropies are not correlation measures. 

Nevertheless, the CCNR negativity provides a useful entanglement witness, which is computable for quantum many-body states and exhibits new universal features~\cite{Yin:2022toc}. Moreover, the operator entanglement entropy is a key quantifier of the complexity of a reduced density matrix, notably in out-of-equilibrium situations, and it can be measured experimentally \cite{Rath:2022qif}.

\section{Reflected entropies and CCNR negativity for free theories}\lb{Gauss_states}

In this section, we compute the $(m,n)$--R\'enyi reflected entropies and CCNR negativities for Gaussian (free) fermionic and bosonic systems in any dimension from the two-point correlation functions of the theory. Doing so, we generalize the results of \cite{Bueno:2020vnx,Bueno:2020fle} for the $m=1$ reflected entropy to arbitrary $m$.

\subsection{Free fermions}\label{sec:FF}

Consider the (diagonalized) fermionic Gaussian state
\bea
\rho_{AB} &= \bigotimes_k \frac{e^{-\eps_k c_k^\dagger c_k}}{1+e^{-\eps_k}}\nn\\
&=\bigotimes_k \big[(1-n_k)|0_k\ra\la0_k| + n_k |1_k\ra\la1_k| \big]\spp,
\eea
where $e^{-\eps_k}=n_k/(1-n_k)$ with $n_k$ being the occupation number at a given wave vector $k$, and the $c_k$’s are fermionic operators satisfying $\{c_i,c_j^\dagger\}=\delta_{ij}$.
The generalized purification $|\Omega_m\ra$, see \eqref{Choi}, can be constructed as
\bea
|\Omega_m\ra 
 \hspace{-1pt}=\hspace{-1pt}\frac{1}{\sqrt{\Tr \rho_{AB}^m}} \bigotimes_k\hspace{-1pt}\big[(1-n_k)^{m/2} + n_k^{m/2}c_k^\dagger \tilde{c}_k^\dagger  \big]|0_k\ra|\tilde{0}_k\ra,
\eea
where $\tilde{c}_k$ are copies of $c_k$ introduced in the vectorization process. The correlation matrix of the state $|\Omega_m\ra$ reads
\bea
&{\bf C}_{kk'}^{(m)} = \la\Omega_m| \begin{pmatrix}
c_k^\dagger\\\
\tilde{c}_{k}
\end{pmatrix}\hspace{-2pt}\begin{pmatrix}
c_{k'} & \tilde{c}_{k'}^\dagger\\
\end{pmatrix} |\Omega_m\ra\spp\nn\\
& =\frac{\delta_{kk'}}{n_k^m +(1-n_k)^m} \hspace{-2pt}
\begin{pmatrix}
n_k^m & (n_k(1-n_k))^{m/2}\\
(n_k(1-n_k))^{m/2} \hspace{-3pt} & (1-n_k)^m
\end{pmatrix} \hspace{-2pt}.
\eea
Performing a Fourier transform leads to the correlation matrix in the spatial basis
\bea\lb{correl_FF}
{\bf C}^{(m)}=\spm
\begin{pmatrix}
\displaystyle\frac{C_{AB}^m}{C_{AB}^m+(1-C_{AB})^m} & \displaystyle\frac{\big(C_{AB}(1-C_{AB})\big)^{m/2}}{C_{AB}^m+(1-C_{AB})^m}\\
\displaystyle\frac{\big(C_{AB}(1-C_{AB})\big)^{m/2}}{C_{AB}^m+(1-C_{AB})^m} & \displaystyle\frac{(1-C_{AB})^m}{C_{AB}^m+(1-C_{AB})^m}
\end{pmatrix}\spm\spm,
\eea
where $C_{AB}$ is the two-point correlation matrix for $\rho_{AB}$. Introducing ${\bf C}_A^{(m)}$ as the restriction of ${\bf C}^{(m)}$ to subsystem $A$, the R\'enyi reflected entropy $S^R_{m,n}$ is obtained as
\bea
S^R_{m,n} = \frac{1}{1-n}\Tr\log\left[\big({\bf C}^{(m)}_A\big)^n + \big(1-{\bf C}^{(m)}_A\big)^n \right].
\eea
For $m=1$, this formula reduces to that for the $n$-R\'enyi reflected entropy derived in \cite{Bueno:2020vnx}, while we recover the operator entanglement entropy \cite{Rath:2022qif} for $m=2$. 
The $(m,n)$--R\'enyi CCNR is obtained using \eqref{CCNR_RE}, explicitly 
\bea
\E_{m,n} &= \Tr\log\left[\big({\bf C}^{(m)}_A\big)^n + \big(1-{\bf C}^{(m)}_A\big)^n \right] \nn\\
& \qquad\quad + n \spp\Tr\log\left[C_{AB}^m + (1-C_{AB})^m\right].
\eea
For $n=1$, we recover $\E_{m,1}=\log\Tr \rho_{AB}^m$. Moreover, for $\rho_{AB}$ pure we get $\E_{m,n}=2\log\Tr\spp\rho_A^n$, in agreement with~\eqref{eq:ccnr_pure}. Indeed, in that case we have $C_{AB}^2=C_{AB}$ such that ${\bf C}^{(m)}_A= C_{A}\oplus (1-C_{A})$, where $C_A$ is the correlation matrix $C_{AB}$ restricted to region $A$.

\subsection{Free bosons}

For Gaussian scalar theories invariant under time reflection, the relevant two-point functions to consider are that of the scalar and the conjugate momentum. Let $\phi_i$ and $\pi_i$, $i=1,\cdots,N$, be a system of scalars and conjugate momenta acting on the Hilbert space $\HA\otimes\HB$. These Hermitian operators satisfy the canonical commutation relations $[\phi_i,\pi_j]=i\delta_{ij}$ and $[\phi_i,\phi_j]=[\pi_i,\pi_j]=0$. 

Given a density matrix $\rho_{AB}$ acting on $\HA\otimes\HB$, a purification $|\Omega_m\ra$ of $\rho_{AB}^m$ can be constructed using the Choi-Jamiolkowski isomorphism by doubling the original Hilbert space of $\rho_{AB}$ and extending accordingly the bosonic algebra with additional operators $\tilde{\phi}_i$ and $\tilde{\pi}_i$ acting on the second copy of $\HA\otimes\HB$. We refer the reader to \cite{Bueno:2020fle} for details on the doubling of Hilbert space for scalar fields. As conveniently done in \cite{Bueno:2020fle}, let us define $\Psi_i^0 \equiv \phi_i$, $\Psi^1_i \equiv \pi_i$, and similarly for $\tilde{\Psi}^a_i$, $a = 0,1$. We find that the correlators for $|\Omega_m\ra$ read
\bea
&\la\Omega_m|\Psi_{i_1}^{a_1}\cdots \Psi_{i_p}^{a_p} \tilde{\Psi}_{j_1}^{b_1}\cdots \tilde{\Psi}_{j_q}^{b_q}|\Omega_m\ra \nn\\\lb{correl_doubledH}
& \qquad =\frac{(-1)^{\sum_q \spm b_q}}{\Tr \rho_{AB}^m} \spp\Tr\Big(\rho_{AB}^{m/2} \Psi_{i_1}^{a_1}\cdots \Psi_{i_p}^{a_p}\rho_{AB}^{m/2}\tilde{\Psi}_{j_q}^{b_q}\cdots \tilde{\Psi}_{j_1}^{b_1}\Big).
\eea
Organizing the scalars in a single field $\Phi$ where $\Phi_i \equiv \phi_i$ and $\Phi_{i+N}\equiv \tilde{\phi}_i$ with $i = 1,\cdots, N$, and similarly for the momenta $\Pi_i$, we are interested in the following two-point correlation functions: 
\bea
\hspace{-3pt}{\bf\Phi}_{kl}^{(m)}\spm\spm =\la\Omega_m| \Phi_k\Phi_l|\Omega_m\ra\spp, \quad {\bf\Pi}_{kl}^{(m)}\spm\spm =\la\Omega_m| \Pi_k\Pi_l|\Omega_m\ra\spp,
\eea
with $k,l = 1,\cdots, 2N$. These two-point functions have been computed in \cite{Bueno:2020fle} using \eqref{correl_doubledH} derived for $m=1$. They depend on the density matrix $\rho_{AB}$ for the first $N$ scalars only. 
Performing the traces, we obtain  
\bea\lb{correl_FB_phi_f}
&{\bf \Phi}^{(m)}=
\begin{pmatrix}
\displaystyle f_m(XP) \spp X & \displaystyle g_m(XP)\spp X \vspace{5pt}\\
\displaystyle g_m(XP)\spp X & \displaystyle f_m(XP)\spp X
\end{pmatrix}\spm\spm,\\\lb{correl_FB_pi}
&{\bf \Pi}^{(m)}=
\begin{pmatrix}
\displaystyle P\spp f_m(XP) & \displaystyle - P\spp g_m(XP) \vspace{5pt}\\
\displaystyle -P\spp g_m(XP) & \displaystyle P\spp f_m(XP)
\end{pmatrix}\spm\spm,
\eea
where we defined
\bea
&f_m(M)=\frac12\frac{\big(\sqrt{M}+1/2\big)^m+\big(\sqrt{M}-1/2\big)^m}{\big(\sqrt{M}+1/2\big)^m-\big(\sqrt{M}-1/2\big)^m}\sqrt{M}^{-1}\spp,\\
&g_m(M)=\frac{\big(M-1/4\big)^{m/2}}{\big(\sqrt{M}+1/2\big)^m-\big(\sqrt{M}-1/2\big)^m}\sqrt{M}^{-1}\spp,\nn
\eea
and $X_{ij}\spm\spm\spm=\spm\spm\Tr(\rho_{AB}\phi_i\phi_j)$ and $P_{ij}\spm\spm=\spm\spm\Tr(\rho_{AB}\pi_i\pi_j)$ are the usual two-point correlation functions. The $(m,n)$--R\'enyi reflected entropy follows from the correlation matrix ${\bf C}_A^{(m)}=\sqrt{{\bf \Phi}^{(m)}_A {\bf \Pi}^{(m)}_A}$, i.e.
\bea
S^R_{m,n} = \frac{1}{n-1}\Tr\log\left[\big({\bf C}^{(m)}_A+1/2\big)^n - \big({\bf C}^{(m)}_A-1/2\big)^n \right]\spm.
\eea
For $m=1$, this formula reduces to that for the $n$-R\'enyi reflected entropy derived in \cite{Bueno:2020fle}. For $m=2$, we have the operator entanglement entropy which was considered for free fermions only in \cite{Rath:2022qif}. 
The $(m,n)$--R\'enyi CCNR is obtained using \eqref{CCNR_RE}, explicitly 
\bea
\E_{m,n} &= - \Tr\log\left[\big({\bf C}^{(m)}_A+1/2\big)^n - \big({\bf C}^{(m)}_A-1/2\big)^n \right] \nn\\
& \qquad\; - n \spp\Tr\log\left[\big(C_{AB}+1/2\big)^m - \big(C_{AB}-1/2\big)^m\right],
\eea
where $C_{AB}=\sqrt{XP}$.
For $\rho_{AB}$ pure, we recover $\E_{m,n}=2\log\Tr\spp\rho_A^n$ since $C_{AB}=1/2$ such that ${\bf C}^{(m)}_A= C_{A}\oplus C_{A}$, with $C_A=\sqrt{X_AP_A}$.

\section{Symmetry resolution}\lb{sec:sym-res}
In this section we investigate the symmetry resolution of the reflected entropies and the CCNR negativity in the context of quantum many-body systems with a global conserved $U(1)$ charge $Q$. 

\subsection{Definitions}

Let us consider a state $\rho_{AB}$ which commutes with a global charge $Q=Q_A+Q_B$, and we further assume $Q=Q^T$. Here, $Q_A$ and $Q_B$ are the local charges for each subsystem. As we show in Appendix~\ref{sec:QImb} (see also \cite{Milekhin:2022zsy,Rath:2022qif}), the reflected density matrix $\rho_{AA}^{(m)}$ commutes with the charge imbalance operators of the two copies of $\HA$,
\begin{equation}
\label{eq:commutation_CI}
   [\Q_A,\rho_{AA}^{(m)}]=0,
\end{equation}
where 
\begin{equation}
    \Q_A = Q_A \otimes \mathbb{I}_{A}-\mathbb{I}_{A}\otimes Q_A \spp.
\end{equation}

The commutation relation \eqref{eq:commutation_CI} implies that $\rho_{AA}^{(m)}$ has a block-diagonal structure where the blocks pertain to different eigenvalues $q$ of $\Q_A$. We have
\begin{equation}
    \rho_{AA}^{(m)} = \bigoplus_q \Big(p_q^{(m)}\rho_{AA}^{(m)}(q)\Big)
\end{equation}
with $\sum_q p_q^{(m)}=1$, and $\Tr \big( \rho_{AA}^{(m)}(q)\big)=1$.
Here $p_q^{(m)}$ is the probability of measuring the eigenvalue~$q$ of the charge $\Q_A$ in the state $\rho_{AA}^{(m)}$, and $\rho_{AA}^{(m)}(q)$ is the reflected density matrix of the charge sector $q$.

The symmetry-resolved $(m,n)$--R\'enyi reflected entropies are defined as 
\bea\lb{REmnq}
S^{R}_{m,n}(q)=\frac{1}{1-n}\log\Tr\big(\rho_{AA}^{(m)}(q)\big)^n\, .
\eea

\pagebreak
\noindent Using \eqref{eq:En_SnR}, the symmetry-resolved operator entanglement entropies read 
\begin{equation}
    E_n(q) = S^{R}_{2,n}(q),
\end{equation}
and this definition corresponds to that introduced in~\cite{Rath:2022qif}.

Similarly as for the standard entanglement entropy, there is a simple relation between the total reflected entanglement entropy and the symmetry-resolved ones, 
\begin{equation}
    S^R_{m,1}= \sum_q p_q^{(m)} S^R_{m,1}(q) - \sum_q p_q^{(m)} \log p_q^{(m)}.
\end{equation}
By analogy with entanglement entropy \cite{exp-lukin}, we call the first term the configurational reflected entropy, and the second one the number reflected entropy. There is a similar, but more cumbersome, relation for reflected entropies with R\'enyi index $n \neq 1$ that we do not reproduce here.

Combining the relation \eqref{eq:rhoAAmRR} and the commutation relation \eqref{eq:commutation_CI}, it follows that $ [\Q_A,R_mR_m^\dagger]=0$, and hence $R_mR_m^\dagger$ also has a block-diagonal structure. However, since the matrix $R_mR_m^\dagger$ does not have unit trace, there is no natural (normalized) definition for the symmetry-resolved $(m,n)$--CCNR negativity. By analogy with \eqref{CCNR_RE}, we choose to define it as 
\begin{equation}\label{eq:EmnqDef}
\begin{split}
    \E_{m,n}(q) &= (1-n)S^R_{m,n}(q) + n (1-m) S_m(\rho_{AB}) \,, 
    \end{split}
\end{equation}
and the symmetry-resolved CCNR negativity is obtained by setting $m=2$ and taking the limit $n\to 1/2$. The relation between total and symmetry-resolved CCNR negativities reads
\begin{equation}
    \eE^{\E_{m,n}} = \sum_q (p_q^{(m)})^n\,\eE^{\E_{m,n}(q)} ,
\end{equation}
similarly to the charge-imbalance-resolved PPT logarithmic negativity \cite{cgs-18,mbc-21,PBC22}.

\subsection{Charged moments}

Extracting the contribution of each charge sector to the total entanglement is a challenging task, since, in principle, it requires the knowledge of the total density matrix. Fortunately, a way to circumvent this issue was proposed in \cite{GS18}, where the idea is to compute so-called charged moments and their Fourier transform. We adapt these ideas to the reflected density matrix, and define the corresponding charged moments as
\begin{equation}
    Z^R_{m,n}(\alpha) = \Tr\left(\big(\rho_{AA}^{(m)}\big)^n \eE^{\ir \alpha \Q_A}\right)\spm.
\end{equation}
Taking the Fourier transform yields the symmetry-resolved moments,
\begin{equation}
    \Z^R_{m,n}(q) = \int_{-\pi}^{\pi}\frac{\dd \alpha}{2\pi} \eE^{-\ir q \alpha}Z^R_{m,n}(\alpha)\spp, \quad\; p_q^{(m)} = \Z^R_{m,1}(q)\spp,\;
\end{equation}

\pagebreak
\noindent and the symmetry-resolved R\'enyi reflected entropies are 
\begin{equation}
\label{eq:SRRE}
    S^R_{m,n}(q) = \frac{1}{1-n}\log \left(\frac{ \Z^R_{m,n}(q) }{ (\Z^R_{m,1}(q))^n }\right)\spm.
\end{equation}
Taking $m=2$ in \eqref{eq:SRRE} yields the symmetry-resolved operator entanglement entropies \cite{Rath:2022qif}, while taking the limit $m\rightarrow 1$ yields the symmetry-resolved reflected entropies.

For the CCNR R\'enyi negativity, the natural charged moments to consider are \cite{Milekhin:2022zsy}
\begin{equation}
\begin{split}\label{eq:ZmnDef}
    Z_{m,n}(\alpha) &= \Tr \left( \big( R_mR_m^\dagger \big)^n \eE^{\ir \alpha \Q_A} \right) \\
   & =  Z^R_{m,n}(\alpha) \left(\Tr \rho_{AB}^m\right)^n\spm,
    \end{split}
\end{equation}
and their Fourier transform yields the symmetry-resolved CCNR moments
\begin{equation}
    \Z_{m,n}(q) = \int_{-\pi}^{\pi}\frac{\dd \alpha}{2\pi} \eE^{-\ir q \alpha}Z_{m,n}(\alpha)\spp.
\end{equation}
We have
\begin{equation}
\begin{split}
    \E_{m,n}(q) &=\log \left(\frac{ \Z_{m,n}(q) }{ (p_q^{(m)})^n }\right)\spm  \\[.3cm]
   &= (1-n)S^R_{m,n}(q) + n (1-m) S_m(\rho_{AB}) \,.
    \end{split}
\end{equation}

\subsection{Conformal field theory results}

In the context of one-dimensional quantum critical system described by a CFT, a large body of results regarding entanglement measures can be obtained by means of the replica trick \cite{Calabrese:2004eu,Calabrese:2009qy,Calabrese:2012ew} and twist fields correlation functions \cite{cardy2008form}. Reflected entropies and the CCNR partition functions $Z_{m,n}$, see \eqref{eq:ccnr_moments}, can also be computed using similar methods \cite{Dutta:2019gen,Yin:2022toc}.

We take the regions $A$ and $B$ to be adjacent segments of respective lengths $\ell_a$ and $\ell_b$, see Fig.\,\ref{fig_tri}. Introducing the relevant twist operator conformal weights~\cite{cardy2008form,Dutta:2019gen}
\begin{equation}
\begin{split}
    &h_n = \frac{c}{24}\left( n-\frac 1n \right), \\
    &h'_{m,n} = n h_m =  n\frac{c}{24}\left( m-\frac 1m \right),
\end{split}
\end{equation}
where $c$ is the central charge of the CFT, the partition function $Z_{2,n}$ for two adjacent intervals can be expressed as \cite{Yin:2022toc}
\begin{equation}\lb{CCNR_moments_adj_2}
    Z_{2,n} \propto (\ell_a \ell_b)^{-4 h_n} (\ell_a+\ell_b)^{4 h_n-4 h'_{2,n}}\spp,
\end{equation}
and for the more generic reflected case we find
\begin{equation}\lb{CCNR_moments_adj}
    Z_{m,n} \propto (\ell_a \ell_b)^{-4 h_n} (\ell_a+\ell_b)^{4 h_n-4 h'_{m,n}}\spp.
\end{equation}

To investigate charged CCNR moments, we follow the approach developed in the  context of charged \cite{belin2013holographic,dowker2016conformal,dowker2017conformal} and symmetry-resolved R\'enyi entropies \cite{GS18,bons,Murciano:2020vgh}, where charged moments correspond to partition functions on multisheeted Riemann surfaces with twisted boundary conditions, such that the total phase accumulated upon going through the entire surface is $\alpha$. Entanglement-related quantities are then obtained as correlation functions of modified twist fields. In particular, the partition function $Z_{2,n}(\alpha)$ corresponds to a correlation function of modified twist fields in $2n$ copies of the original theory. These are operators obtained by the fusion of standard twist fields of weight $h_n$ with operators generating the flux $\alpha$, with weight $h_\alpha$.   The resulting composite twist fields behave as primaries with weight 
\begin{equation}
   h_{n}(\alpha) = h_n+\frac{h_\alpha}{n}.
\end{equation}
As an example, for the ground state of a Luttinger liquid with parameter $K$, the weight $h_\alpha$ reads \cite{GS18,bons}
\begin{equation}
     h_\alpha = \frac{K}{2}\left(\frac{\alpha}{2\pi}\right)^2 \spp.
\end{equation}

To maintain charge neutrality in the correlation function corresponding to $Z_{2,n}(\alpha)$, fields with charge $\alpha$ are inserted on odd copies, and fields with opposite charge are inserted on even copies. In the case of adjacent intervals, twist fields at the ends of the interval $A \cup B$ connect copies of different parities, and therefore fields with opposite charge cancel. However, at the interface between $A$ and $B$, odd and even copies are decoupled \cite{Yin:2022toc} and hence the only charge dependence arises from that part. Generalizing the argument to arbitrary reflected index~$m$, we find
\begin{equation}\lb{CCNR_moments_adj_alpha}
    Z_{m,n}(\alpha) \propto \left(\frac{\ell_a \ell_b}{\ell_a+\ell_b}\right)^{-4 h_{n}(\alpha)} (\ell_a+\ell_b)^{-4 n h_{m}(0)}\spp,
\end{equation}
where we assumed $h_\alpha=\bar{h}_\alpha$ and $h_m(0) = h_m$. We note that the reflected moments $Z^R_{m,n}(\alpha)$ are simply obtained by setting $h'_{m,n}\to0$ in \eqref{CCNR_moments_adj}, and the charged reflected moments are by retaining the first factor only in \eqref{CCNR_moments_adj_alpha}, 
\begin{equation}\lb{R_moments_adj_alpha}
    Z^R_{m,n}(\alpha) \propto \left(\frac{\ell_a \ell_b}{\ell_a+\ell_b}\right)^{-4 h_{n}(\alpha)}\spp. 
\end{equation}
For $\alpha=0$, we recover the known result for the reflected moments \cite{Dutta:2019gen}. In particular, the reflected entropies do not depend on $m$ at leading order for adjacent regions.

\begin{figure}[t]
\centering
\includegraphics[scale=1]{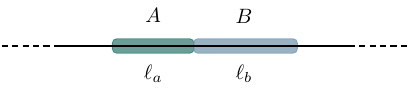}
\caption{Two adjacent intervals $A$ and $B$ of length $\ell_a$ and $\ell_b$, respectively, in an infinite system.}
\lb{fig_tri}
\end{figure}

\pagebreak
\subsection{Symmetry resolution for fermions}

For critical free fermions ($c=1,K=1$) in one dimension, we have~\cite{GS18,bons}
\begin{equation}
    h_\alpha = \frac{1}{2}\left(\frac{\alpha}{2\pi}\right)^2 \spp,
\end{equation}
finding 
\bea\lb{CCNR_moments_adj_fermions}
    &Z_{m,n}(\alpha) \propto \left(\frac{\ell_a \ell_b}{\ell_a+\ell_b}\right)^{-4 h_n^{\rm f}(\alpha)}\left(\ell_a+\ell_b\right)^{-4nh_m^{\rm f}(0)}\spp
\eea
with 
\begin{equation}\label{eq:hnf}
    h_n^{\rm f}(\alpha) = \frac{1}{24}\left( n-\frac 1n \right)+\frac{1}{2n}\left(\frac{\alpha}{2\pi}\right)^2.
\end{equation}

To check our CFT predictions, we consider the one-dimensional tight-biding model with Hamiltonian 
\begin{equation}
    H = -\sum_{j=1}^L (c_{j+1}^\dagger c_j + \textrm{H.c.})\spp.
\end{equation}
Here, $c_j^\dagger,c_j$ are fermion creation and annihilation operators, $L$ is the length of the chain and we assume periodic boundary conditions. This model has a $U(1)$ symmetry generated by the charge operator $Q=\sum_j c_j^\dagger c_j$, and its diagonalization is standard. The ground-state two-point correlation functions in the large-$L$ limit reads
\begin{equation}\label{eq:CFF}
    C_{ij} = \frac{\sin({\frac{\pi}{2}(i-j))}}{\pi (i-j)}\spp.
\end{equation}

Using results from Sec.~\ref{sec:FF} and standard free-fermion techniques \cite{peschel2003calculation,peschel2009reduced, GS18}, we express the the charged moments $Z_{m,n}(\alpha)$ defined in \eqref{eq:ZmnDef} in terms of the two-point correlation matrix as 
\begin{multline}
\log Z_{m,n}(\alpha) = \Tr\log\left[\big({\bf C}^{(m)}_A\big)^n \eE^{\ir \alpha} + \big(1-{\bf C}^{(m)}_A\big)^n \right] 
\\ + n \spp\Tr\log\left[C_{AB}^m + (1-C_{AB})^m\right].
\end{multline}
We note that the logarithmic charged moments have a trivial imaginary part, $\textrm{Im}(\log Z_{m,n}(\alpha)) = \alpha \ell_a$. We thus redefine the fermionic charged moments as
\begin{equation}
     Z_{m,n}(\alpha) \rightarrow \eE^{-\ir \alpha \ell_a} Z_{m,n}(\alpha)
\end{equation}
so that $\log Z_{m,n}(\alpha)$ is real. This redefinition is merely a convention which sets the average value of the charge imbalance to zero.

With all this in hand, we now investigate the charged CCNR moments $Z_{m,n}(\alpha)$ for adjacent intervals of equal length $\ell$ embedded in an infinite chain. In the limit of large $\ell$ we expect 
\begin{equation}
    \log Z_{m,n}(\alpha) =  -4\big(h_n^{\rm f}(\alpha)+nh_m^{\rm f}(0)\big) \log \ell + \cdots 
\end{equation}
where $h_n^{\rm f}(\alpha)$ is given in \eqref{eq:hnf}. We compare this CFT prediction with exact numerical diagonalization in Fig.~\ref{fig_fermions} and find a perfect agreement. Each numerical point is obtained by fitting the leading coefficient of $ \log Z_{m,n}(\alpha)$ for increasing values of $\ell$ and fixed $m,n,\alpha$.

\begin{figure}[t]
\centering
\includegraphics[scale=1.01]{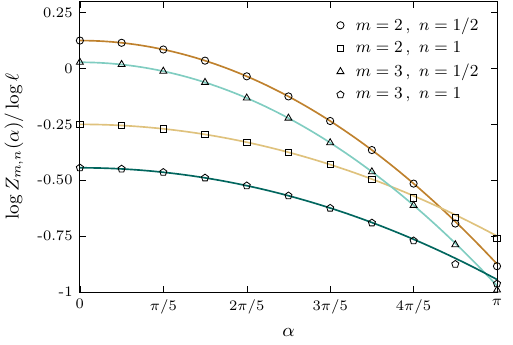}
\caption{Charged CCNR moments for free critical fermions on the infinite chain for two adjacent intervals of same length~$\ell$. The solid lines correspond to the analytical prediction $-4(h_n^{\rm f}(\alpha)+nh_m^{\rm f}(0))$  where $h_n^{\rm f}(\alpha)$ is given in \eqref{eq:hnf}.}
\lb{fig_fermions}
\end{figure}

Computing the Fourier transform of \eqref{CCNR_moments_adj_fermions} yields the symmetry-resolved CCNR and reflected moments as
\begin{equation}\label{eq:ZmnqR}
    \Z^{(R)}_{m,n}(q) = Z^{(R)}_{m,n}(0) \sqrt{\frac{n \pi}{2\log(\frac{\ell_a\ell_b}{\ell_a+\ell_b})}} \exp \left(-\frac{n \pi^2 q^2}{2 \log(\frac{\ell_a\ell_b}{\ell_a+\ell_b})}\right)\spm,
\end{equation}
valid both for $\Z_{m,n}(q)$ and $\Z^R_{m,n}(q)$, since these charged moments have same $\alpha$ dependence. Expression \eqref{eq:ZmnqR} being exact at leading order, we do not check it numerically.

As both symmetry-resolved CCNR negativity $\E_{m,n}(q)$ and reflected entropies $S^R_{m,n}(q)$ have the same charge dependence, see \eqref{eq:EmnqDef}, we focus our discussion on the latter. Combining \eqref{eq:SRRE} and \eqref{eq:ZmnqR}, we find
\bea\label{eq:EQSmn}
S^R_{m,n}(q) = S^R_{m,n} &-\frac12\log\log\hspace{-1pt}\left(\frac{\ell_a \ell_b}{\ell_a+\ell_b}\right)\nn\\
&\qquad +\frac12\log \frac{\pi n^{1/(1-n)}}2 +\cdots\spp,
\eea
where $S^R_{m,n}$ is the total $(m,n)$--R\'enyi reflected entropy. This result implies that, at leading order, the symmetry-resolved reflected entropies do not depend on the charge sector. This phenomena has been observed for symmetry-resolved entanglement entropies in numerous physical situations, and has been dubbed \textit{equipartition of entanglement} \cite{equi-sierra}. Of course, \eqref{eq:EQSmn} is a CFT prediction and thus holds in the large-$\ell$ limit. At finite size, we expect a subleading $q$-dependent correction. For the entanglement entropies this correction is of order $(q/\log \ell )^2$, see \cite{bons}, and we expect a similar behavior for the reflected entropies. It is an important open problem to perform exact finite-size calculation of symmetry-resolved reflected entropies to determine the charge-dependent corrections. We leave this issue for further investigations.

\subsection{Symmetry resolution for complex bosons}

Consider a complex scalar field $\varphi$ of mass $\mu$ with Lagrangian density $\mathcal{L}=\partial_\nu\varphi^\dagger\partial^\nu\varphi-\mu^2\varphi^\dagger\varphi$. This Lagrangian exhibits a $U(1)$ symmetry, i.e.~it is invariant under phase transformations of the field $\varphi\rightarrow e^{\ir \theta}\varphi$. 
The complex scalar field can be written in terms of two real ones defining $\varphi = (\varphi_1 + \ir \varphi_2)/\sqrt{2}$. Its Hamiltonian $H$ is the sum of two identical Hamiltonians corresponding to each real scalars, $H=\sum_{i=1,2}\frac12\int dx \big(\pi_i^2 + (\partial_x\varphi_i)^2 + \mu^2\varphi_i^2\big)$.
Introducing the creation and annihilation operators $a_i^\dagger(p), \, a_i(p)$ with momentum $p$ for each real scalars $i = 1,2$, the Hamiltonian and the conserved charge $Q$ associated to the $U(1)$ symmetry can be written in terms of particles and antiparticles modes operators $a_p = (a_1(p) + \ir a_2(p))/\sqrt{2}$ and $b_p = (a_1^\dagger(p) + \ir a_2^\dagger(p))/\sqrt{2}$\,:
\bea
&H=\int\frac{dp}{2\pi}\omega_p\big(a_p^\dagger a_p + b_p^\dagger b_p \big)\spp,\lb{Ham_Cscal}\\
&Q=\int\frac{dp}{2\pi}\big(a_p^\dagger a_p - b_p^\dagger b_p \big)\spp,
\eea
where $\omega_p^2=\mu^2+p^2$. In real space, the value of the conserved charge in a given subsystem $A$ reads
\bea\lb{charge_Cscal}
Q_A=\int_A dx\big(a(x)^\dagger a(x) - b(x)^\dagger b(x) \big)\equiv N_A^a-N_A^b\spp,
\eea
which is the difference between the number of particles and the number of antiparticles.

Using \eqref{charge_Cscal}, the charge imbalance operators $\Q_A$ of the two copies of $\HA$ that commutes with the reflected density matrix can be written as $\Q_A=\N_A^a-\N_A^b$ where
\bea\lb{charge_imb_Cscal}
\N_A^a = N_A^a \otimes \mathbb{I}_{A}-\mathbb{I}_{A}\otimes N_A^a\spp,
\eea
and similarly for $\mathcal{N}_A^b$. Since the Hamiltonian of a complex scalar field is identical to that of two decoupled real scalar fields, the density matrix factorizes accordingly. The charged moments of the reflected density matrix for the complex scalar field can be written in terms of those for a single scalar field, i.e.
\bea
Z^R_{m,n}(\alpha) = \Big| \Tr\spm\left(\spm\big(\rho_{AA}^{(m)}\big)^n \eE^{\ir \alpha \N_A}\spm\right)\spm\spm\Big|^2\spm,
\eea
where we used \eqref{charge_imb_Cscal}, and $\rho_{AA}^{(m)}$ corresponds to the reflected density matrix for a single scalar field.

On the lattice, the (complex) scalar field is realized by the (complex) harmonic chain. Using the results of Sec.~\ref{Gauss_states} for bosonic states, the charged moments for the complex harmonic chain are obtained as
\bea
Z^R_{m,n}(\alpha) = \det\Big| \big({\bf C}^{(m)}_A+1/2\big)^n - \eE^{\ir \alpha}\big({\bf C}^{(m)}_A-1/2\big)^n \Big|^{-2}\spp,
\eea
where the correlation matrix ${\bf C}^{(m)}_A$ is defined in terms of the two-point functions of the real harmonic chain.

Let us now focus on adjacent subsystems $A$ and $B$, of respective lengths $\ell_a$ and $\ell_b$. In the critical (massless) limit $\mu\rightarrow 0$, the charged reflected moments take the same form as in CFT, i.e.
\bea\lb{reflected_moments_adj_boson}
    &Z_{m,n}^R(\alpha) \propto \left(\frac{\ell_a \ell_b}{\ell_a+\ell_b}\right)^{-4 h_n^{\rm b}(\alpha)}\spp,
\eea
only with a nonstandard weight $h_n^{\rm b}(\alpha)$, where a linear term in $\alpha$ is present (see, e.g., \cite{Murciano:2019wdl,Murciano:2020vgh} for similar findings concerning the charged entanglement entropies),
\bea\lb{weight_boson}
h_n^{\rm b}(\alpha) = \frac{1}{12}\left( n-\frac 1n \right)+\frac{|\alpha|}{4\pi n}-\frac{1}{2n}\spm\left(\frac{\alpha}{2\pi}\right)^2\spm.
\eea
Similarly, the charged CCNR moments are given by
\bea\lb{CCNR_moments_adj_boson}
    &Z_{m,n}(\alpha) \propto \left(\frac{\ell_a \ell_b}{\ell_a+\ell_b}\right)^{-4 h_n^{\rm b}(\alpha)}\left(\ell_a+\ell_b\right)^{-4nh_m^{\rm b}(0)}\spp.
\eea

As a check of our formulas, we have numerically computed the charged CCNR moments for the complex boson in the critical regime $\mu\ll1$. We take two adjacent regions of same length $\ell$ on the infinite chain. From \eqref{CCNR_moments_adj_boson}, we expect $\log Z_{m,n}(\alpha) = -4(h_n^{\rm b}(\alpha)+nh_m^{\rm b}(0)) \log \ell + \cdots$. We report in Fig.\,\ref{fig_bosons} our numerical results for the leading logarithmic coefficient of $\log Z_{m,n}(\alpha)$ as a function of $\alpha$ for different $m$ and $n$. We observe a perfect agreement between \eqref{CCNR_moments_adj_boson} and the numerics.

\begin{figure}[t]
\centering
\includegraphics[scale=1.01]{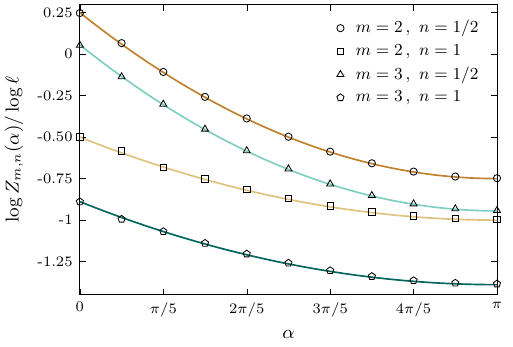}
\caption{Charged CCNR moments for the free complex boson on the line in the critical regime $\mu\spm\ll\spm1$ for two adjacent intervals of same length $\ell$. The solid lines correspond to the CFT prediction $-4(h_n^{\rm b}(\alpha)+nh_m^{\rm b}(0))$, where $h_n^{\rm b}(\alpha)$ is given in \eqref{weight_boson}.}
\lb{fig_bosons}
\end{figure}

We are interested in the Fourier transform of \eqref{reflected_moments_adj_boson}. In the saddle point approximation, the term in $\alpha^2$ in \eqref{weight_boson} can be neglected, yielding
\bea\label{eq:ZmnRqBosons}
\Z^R_{m,n}(q)=Z^R_{m,n}(0)\frac{n \log(\frac{\ell_a \ell_b}{\ell_a+\ell_b})}{n^2\pi^2q^2+\log^2(\frac{\ell_a \ell_b}{\ell_a+\ell_b})}\spp.
\eea
Similarly as for free fermions, we only investigate the symmetry-resolved reflected entropies since they have the same charge dependence as the symmetry-resolved CCNR negativities. Combining \eqref{eq:SRRE} and \eqref{eq:ZmnRqBosons} we find
\begin{multline}
    S^R_{m,n}(q) =  S^R_{m,n} - \log\log\hspace{-1pt}\left(\frac{\ell_a \ell_b}{\ell_a+\ell_b}\right)\\+\frac{\log n}{1-n} + \frac{n\pi^2 q^2}{\log^2(\frac{\ell_a \ell_b}{\ell_a+\ell_b})}+\cdots\spp,
\end{multline}
where the subleading terms are of order $\big(\hspace{-1pt}\log\ell\big)^{-3}$. We thus observe an equipartition of the reflected entropies for free bosons. However, as opposed to the free-fermion case, the CFT prediction yields the first charge-dependent subleading correction, which is of order $\big(q/\log\ell\big)^{2}$. These results are very similar to those obtained for the standard symmetry-resolved entropies for free bosons~\cite{Murciano:2019wdl,Murciano:2020vgh}.

\section{Conclusion}\lb{conclu}

In this paper, we have investigated the $(m,n)$--R\'enyi reflected entropies and CCNR negativities for CFTs and free theories. We first defined a reflected version of the CCNR negativity, denoted $\E_{m,n}$, and discussed its connections with other quantities of interest, namely the reflected entropies and the operator entanglement entropies. On general ground, we have shown that the CCNR negativity and operator entanglement entropy are not monotonically nonincreasing under partial trace; hence, they are not correlation measures for mixed state. Nevertheless, these quantities present interesting universal properties.
We have derived general formulas for the $(m,n)$--R\'enyi reflected entropies and CCNR negativities, valid for free fermionic and scalar fields in arbitrary dimensions. The resulting expressions are fully determined in terms of the fields' correlators, making them suitable for lattice calculations. 

The main part of this work concerned the symmetry resolution of reflected entropies and CCNR negativities in the context of quantum many-body systems with a global $U(1)$ conserved charge. For two adjacent regions in an infinite system described by a $(1+1)$-dimensional CFT, we have obtained the charged reflected moments and CCNR moments, see \eqref{CCNR_moments_adj_alpha} and \eqref{R_moments_adj_alpha}, focusing then on the former as they share the same charge dependence. Since the exact form of these charged moments depends on the details of the underlaying theory, we have considered both a free fermionic model and a free bosonic model. Using our general formulas for the reflected entropies and CCNR negativities in free theories, we have tested our CFT predictions, finding perfect agreement in both cases. Our numerical investigations also confirm the CFT results of \cite{Yin:2022toc} for the (uncharged) CCNR negativity, for two new examples--free fermions and bosons--which were not considered in \cite{Yin:2022toc}. For free fermions, the symmetry-resolved reflected entropies do not depend on the charge sector up to subleading terms, which we conjecture to be of order $\big(q/\log\ell\big)^{2}$, thereby implying equipartition, similarly to the entanglement entropy. For free bosons, the CFT prediction yields the first charge-dependent subleading correction, which is of order $\big(q/\log\ell\big)^{2}$.
Note that we worked with infinite systems for simplicity, but our results can easily be generalized to periodic systems with finite size $L$ by replacing each subregion length $\ell_i$ by $\frac L\pi\sin\frac{\pi\ell_i}{L}$.

There are several future avenues worth exploring. An important direction would be to study the symmetry resolution of reflected entropies and CCNR negativities for disjoint subsystems. The CFT calculation in this case is more involved than for adjacent regions, and we expect would yield new insights on multipartite entanglement. One could also generalize our findings to finite temperature states, and to quench dynamics. Note that the symmetry-resolved operator entanglement entropy was investigated in \cite{Rath:2022qif} in out-of-equilibrium situations, and we expect similar results for the CCNR negativities and reflected entropies, in particular the presence of an entanglement barrier. It would be interesting to derive new exact lattice results for the reflected entropies and CCNR negativities, both in and out of equilibrium. In higher dimensions, the study of skeletal regions \cite{Berthiere:2021nkv}, i.e.~regions $A,B$ that have no volume, could lead to new analytical results for reflected entropies and CCNR negativities. Finally, it is worth investigating reflected entropies and CCNR negativities for other mixed quantum states, such as Rokhsar-Kivelson states \cite{RK88,PhysRevB.35.8865} whose Hilbert space is spanned by the configurations of an underlying statistical model (see \cite{Boudreault:2021pgj,Parez:2022ind} for recent developments on their entanglement structure).

\begin{acknowledgements}
We warmly thank Sara Murciano, Filiberto Ares and J\'erôme Dubail for useful discussions and sharing some preliminary results. We also thank Pablo Bueno and Riccarda Bonsignori for comments on the manuscript. C.B.~was supported by a CRM-Simons Postdoctoral Fellowship at the Universit\'e de Montr\'eal. G.P.~holds an FRQNT and a CRM-ISM Postdoctoral Fellowship, and acknowledges support from the Mathematical Physics Laboratory of the CRM. 
\end{acknowledgements}

\appendix

\section{Proof of nonmonotonicity of CCNR negativity and operator entanglement entropy}\label{app:nonm}

In this appendix, we show that the CCNR R\'enyi negativity and the R\'enyi operator entanglement entropy are not monotonically nonincreasing under partial trace. To do so, it suffices to find counterexamples to the inequality~\eqref{monotone_ineq}, as was done in \cite{Hayden:2023yij} for the reflected entropy~$S^R_{1,n}$. 

Let us consider the Hilbert space $\HA\otimes\HB\otimes\HC$ of two qutrit $A$ and $B$ and one qubit $C$. 
The counterexample states, labeled by a single parameter $p$, are given by \cite{Hayden:2023yij}
\bea
\rho_{ABC} = \frac{1}{4p+2}\Big(&p\big(|000\ra\la000| +|110\ra\la110| \nn\\
&\;\;+|200\ra\la200|+|210\ra\la210|\big) \nn\\
&\quad\;+|020\ra\la020|+|121\ra\la121| \Big)\spp,
\eea
and we further define $\rho_{AB}=\Tr_{\HC}\rho_{ABC}$. It is straightforward to obtain the reflected densities $\rho_{AA(C)}^{(m)}$ and $\rho_{AA}^{(m)}$ on $\HA\otimes\HA$ from the purifications of $\rho_{ABC}^{m}$ and $\rho_{AB}^{m}$, respectively [see \eqref{Choi}]. They are given by 
\bea
\rho_{AA(C)}^{(m)}&= \frac{1}{4p^m+2}\Big(p^m\big(|00\ra\la00| + 2|22\ra\la22| + |00\ra\la22| \nn\\
&\quad\;+|22\ra\la00|+|11\ra\la11|+|11\ra\la22|+|22\ra\la11|\big) \nn\\
&\qquad+|00\ra\la00|+|11\ra\la11| \Big)\spp,\\
\rho_{AA}^{(m)}&=\rho_{AA(C)}^{(m)} + \frac{1}{4p^m+2}\big(00\ra\la11|+|11\ra\la00| \big)\spp.
\eea
Their corresponding (nonzero) eigenvalues are
\bea
&{\rm Spec}\big(\rho_{AA(C)}^{(m)}\big) \\
&\qquad\quad=\bigg\{\frac{p^m+1}{4 p^m+2},\frac{1+3 p^m\pm\sqrt{9 p^{2 m}-2 p^m+1}}{2 \left(4p^m+2\right)} \bigg\}\spp,\nn\\
&{\rm Spec}\big(\rho_{AA}^{(m)}\big) \\
&\qquad\quad=\bigg\{\frac{p^m}{4 p^m+2},\frac{2+3 p^m\pm\sqrt{9 p^{2 m}-4 p^m+4}}{2 \left(4p^m+2\right)} \bigg\}\spp.\nn
\eea

Using these expressions, one can compute $S^R_{m,n}$ and $\E_{m,n}$ defined in \eqref{REmn} and \eqref{CCNR_RE}, respectively, for any value of $p$ and $m, n$. For large $p$ we find that
\bea\lb{diffRE}
\Delta S^R_{m,n}=
\begin{cases}
\displaystyle\frac{1}{1-n}\frac{1-2^{n}}{1+3^n}\Big(\frac{2}{3}\Big)^n\frac{1}{p^{nm}} + \cdots\spp, & n<1\spp,\vspace{6pt}\\
\displaystyle-\frac{1}{6}\frac{\log p^m}{p^m} + \cdots\spp, & n=1\spp,\vspace{5pt}\\
\displaystyle\frac{1}{1-n}\frac{1-3^{n-2}}{1+3^n}\frac{n}{p^m} + \cdots\spp, & n>1\spp,
\end{cases}
\eea
where $\Delta S^R_{m,n}\equiv S^R_{m,n}(A:B\cup C)-S^R_{m,n}(A:B)$ and the ellipsis denotes terms strictly smaller than $1/p^m$ in the limit $p\rightarrow\infty$. The difference $\Delta S^R_{m,n}$ is thus negative for large $p$ in the range $0<n<2$ and $m>0$. It can be shown to be positive for $n\geqslant 2$ for any $p>0$. This implies that the R\'enyi reflected entropy $S^R_{m,n}$, which reduces to the R\'enyi operator entanglement entropy $E_n$ for $m=2$, is not monotonically nonincreasing under partial trace.

Since $\Tr\rho_{ABC}^m=\Tr\rho_{AB}^m$, we have the difference
\bea
&\E_{n}(A:B\cup C)-\E_{n}(A:B) = \log\frac{\Tr\big(\rho_{AA(C)}^{(2)}\big)^n}{\Tr\big(\rho_{AA}^{(2)}\big)^n}\spp,
\eea
where we set $m=2$. For $n\in(0,1)\cup(1,\infty)$, there exists a critical value $p_0 \simeq 1.284210838876$, solution to the equation $\partial_n\big(\Tr\big(\rho_{AA(C)}^{(2)}\big)^n/\Tr\big(\rho_{AA}^{(2)}\big)^n\big)|_{n=1}=0$, for which we have $\E_{n}(A:B\cup C)-\E_{n}(A:B)<0$, implying that the CCNR R\'enyi negativity is not monotonically nonincreasing under partial trace.

\section{Commutation and charge imbalance}\label{sec:QImb}

In this section, following \cite{Rath:2022qif}, we prove the commutation relation \eqref{eq:commutation_CI} between the reflected density matrix $\rho_{AA}^{(m)}$ and the charge imbalance $\Q_A$.

Let us consider an operator $\Oo$ which acts on $\HA \otimes \HB$ and commutes with the charge $Q=Q_A+Q_B$, namely $[Q,\Oo]=0$. It follows that $\Oo$ takes the form
\begin{equation}
    \Oo = \sum_q \sum_j \lambda_j^{(q)} \Oo_{A,j}^{(q)} \otimes\Oo_{B,j}^{(-q)}\spp,
\end{equation}
where $q$ labels the eigenvalues of $Q$,
\begin{equation}\label{eq:comm_QA_O0}
  \Tr \{(\Oo_{A,j}^{(q)})^\dagger \Oo_{A,j'}^{(q')}\} = \delta_{q,q'}\delta_{j,j'},\quad\;  [Q_A, \Oo_{A,j}^{(q)}] = q \Oo_{A,j}^{(q)}\spp,
\end{equation}
and similarly for $\Oo_{B,j}^{(q)}$. We recast $\Oo$ in components as 
\begin{equation}
    \Oo = \sum_{q,j}\sum_{a,a'}\sum_{b,b'}\lambda_j^{(q)} \langle a|\Oo_{A,j}^{(q)}|a'\rangle\langle b|\Oo_{B,j}^{(-q)}|b'\rangle \, |ab\rangle \langle a'b'|\spp,
\end{equation}
where $\{|a\rangle\}$ and $\{|b\rangle\}$ are orthonormal basis states of $\HA$ and $\HB$. We choose the basis states to be eigenvectors of the charges,
\begin{equation}\label{eq:QAa}
    Q_A |a\rangle = q_a |a\rangle, \quad Q_B |b\rangle = q_b |b\rangle\spp.
\end{equation}
The commutation relation \eqref{eq:comm_QA_O0} together with \eqref{eq:QAa} implies 
\begin{equation}\label{eq:delta_imbalance}
    \langle a|\Oo_{A,j}^{(q)}|a'\rangle \propto \delta_{q_a- q_{a'}, q}, \quad  \langle b|\Oo_{B,j}^{(-q)}|b'\rangle \propto \delta_{q_{b'}- q_{b}, q}\spp.
\end{equation}

The vector version of $\Oo$ reads
\begin{equation}
   | \Oo \rangle = \sum_{q,j}\sum_{a,a'}\sum_{b,b'}\lambda_j^{(q)} \langle a|\Oo_{A,j}^{(q)}|a'\rangle\langle b|\Oo_{B,j}^{(-q)}|b'\rangle\,|ab\rangle \otimes | a'b'\rangle
\end{equation}
and we have
\bea
&\Tr_{\HB\otimes \HB}\big(|\Oo\ra\la\Oo|\big) \nn\\
&\quad=\sum_{q,j}\sum_{\substack{a,a'\\A,A}}(\lambda_j^{(q)})^2  \langle a|\Oo_{A,j}^{(q)}|a'\rangle \langle A'|(\Oo_{A,j}^{(q)})^\dagger|A\rangle \, |aa'\rangle\langle AA'|\spp,
\eea
where we used the orthogonality relation \eqref{eq:comm_QA_O0}. Finally, with \eqref{eq:delta_imbalance} we conclude that each operator $|aa'\rangle\langle AA'|$ in the sum commutes with the charge imbalance 
\begin{equation}
    \Q_A = Q_A \otimes \mathbb{I}_{A}-\mathbb{I}_{A}\otimes Q_A \spp, 
\end{equation}
and therefore 
\begin{equation}
    [\Q_A,  \Tr_{\HB\otimes \HB}\big(|\Oo\ra\la\Oo|\big) ] = 0 \spp.
\end{equation}

\pagebreak
\section{The generalized Markov gap}\lb{apdx:markov}

We consider a generalization of the Markov gap \cite{Hayden:2021gno,Zou:2020bly}, defined as the difference of the $m$-reflected entropy with mutual information~$I$
\bea
M_m \equiv S^R_{m,1} - I\,,
\eea
where we recall that mutual information between two subsystems $A$ and $B$ is given by $I(A:B)= S(\rho_A) + S(\rho_B)-S(\rho_{A\cup B})$. The reflected entropy $m=1$ is lower bounded by mutual information \cite{Dutta:2019gen}, hence the Markov gap is a nonnegative quantity $M\equiv M_1 \ge 0$.
As argued in \cite{Zou:2020bly}, a nonvanishing $M$ indicates irreducible tripartite entanglement.

Using our results of Sec.~\ref{Gauss_states}, we numerically compute $M_m$ for two adjacent subsystems, both for free fermions and free bosons. In the thermodynamic limit, we find
\bea
M_m = 
\begin{cases}
\displaystyle\spp\frac13\log2m \,, \quad &{\rm fermions,}\vspace{6pt}\\ 
\displaystyle\spp\frac13\log2m -\frac12\log m \,, \quad &{\rm bosons.}\\ 
\end{cases}
\eea
For $m=1$, we retrieve the expected universal CFT result \cite{Zou:2020bly} that $\displaystyle M=\frac c3\log2$. However, for $m\neq 1$ we observe discrepancy between fermions and bosons. It would be interesting to understand if $M_m$ is a meaningful quantity in general.

\let\oldaddcontentsline\addcontentsline
\renewcommand{\addcontentsline}[3]{}

\bibliographystyle{utphys} 

\input{CCNR_bib.bbl}

\let\addcontentsline\oldaddcontentsline

\end{document}

%% file: CCNR_bib.bbl
\providecommand{\href}[2]{#2}\begingroup\raggedright\endgroup